\documentclass[11pt,a4paper]{article}
\pdfoutput=1
\usepackage{jheppub}
\usepackage{graphicx,amsmath,mathtools}% Include figure files
\usepackage{epsfig,subfigure}
\usepackage{cancel}

\def\beq{\begin{equation}}
\def\eeq{\end{equation}}
\def\bea{\begin{eqnarray}}
\def\eea{\end{eqnarray}}

\def\3Eqs#1#2#3{Eqs.\ (\ref{#1}), (\ref{#2}) and (\ref{#3})}

\def\to{\longrightarrow}

\def\within#1to#2/{#1\mbox{ to }#2}
\def\lbullet#1,#2,#3,#4/{\Text(#1,#2)[c]{$\bullet$}\Line(#1,#2)(#3,#4)}
\def\textcite#1{Ref.~\cite{#1}}
\def\sL#1{{\tilde L}_{#1}}
\def\sE#1{{\tilde E}_{#1}}
\def\newu{$U(1)_{L_\mu-L_\tau}$}

\def\gino{\tilde{g}}
\def\bino{\tilde{B}}
\def\bpino{\tilde{B^{\prime}}}
\def\wino{\tilde{W}}

\def\hd0ino{\tilde{h^0_{d}}}

\def\hu0ino{\tilde{h^0_{u}}}

\def\sU{{\tilde U}}
\def\sD{{\tilde D}}
\def\sQ{\tilde Q}
\def\sL{\tilde L}
\def\sE{\tilde E}

\def\sneue{{\tilde \nu}_{e}}

\def\sneumu{{\tilde \nu}_{\mu}}

\def\etabar{\bar \eta}
\def\e2spl{\ensuremath{^\clubsuit}}

\def\nn{\nonumber\\*}

\title{Supersymmetric gauged \boldmath U(1)$_{L_{\mu}-L_{\tau}}$ model for electron and muon \boldmath $(g-2)$ anomaly}

\author[a]{Heerak Banerjee}
\author[b]{Bhaskar Dutta}
\author[a]{Sourov Roy}

\affiliation[a]{School of Physical Sciences, Indian Association for the Cultivation of Science,
2A $\&$ 2B Raja S.C. Mullick Road, Kolkata 700 032, India}
\affiliation[b]{Mitchell Institute for Fundamental Physics and Astronomy,
Department of Physics and Astronomy, Texas A\&M University, College Station, TX 77843, USA}
\emailAdd{tphb@iacs.res.in}
\emailAdd{dutta@physics.tamu.edu}
\emailAdd{tpsr@iacs.res.in}
\abstract{Minimal gauged \newu models can provide for an additional source for the muon anomalous magnetic moment however it is difficult to accommodate the discrepancy in the electron magnetic moment in tandem. Owing to the relative sign between the discrepancies in these quantities, it seems unlikely that they arise from the same source. We show that a supersymmetric (SUSY) gauged \newu model can accommodate both the muon and electron anomalous magnetic moments in a very simple and intuitive scenario, without utilizing lepton flavor violation. The currently allowed parameter space in this kind of a scenario is constrained from the latest LHC  and various low energy experimental data,e.g., recent COHERENT data, CCFR, Borexino, BaBaR, supernova etc. These constraints, in conjunction with the requirement to explain both lepton magnetic moments, lead to an upper bound on the first generation slepton mass, a lower bound on the second generation slepton mass and constricts the allowed range for the new gauge boson mass and coupling. The scheme can be probed at the ongoing COHERENT and Coherent CAPTAIN-Mills experiments and at future experiments, e.g., DUNE, BELLE-II etc.}
\begin{document}
\begin{flushright}
MI-TH-2028
\end{flushright}
\maketitle

%%%%%%%%%
\section{Introduction}
The electron and muon magnetic moments are counted amongst one of the most accurately measured quantities today. Their theoretical predictions within the Standard Model(SM) have also been calculated with great precision. Exciting deviations between these quantities have recently incited considerable interest within the fraternity. The measured value of the muon anomalous magnetic dipole moment (MDM) \cite{Bennett:2006fi} stands at $\sim 3.7\sigma$ deviation \cite{Aoyama:2020ynm,Tanabashi:2018oca} from its value predicted from the SM
\begin{equation}
\delta a_\mu = a_{\mu}^{\rm exp} - a_{\mu}^{\rm SM} = 2.79(0.76)\times 10^{-9}.
\end{equation} The ongoing experiment E989 at Fermilab \cite{Grange:2015fou} and the upcoming one at J-PARC \cite{Mibe:2010zz} is expected to achieve a four-fold increase in the accuracy of the experimental measurement. If the central value of the measurement remains unchanged the discrepancy will stand at about $7\sigma$ \cite{Keshavarzi:2018mgv}. Very recently, the $\Delta a_\mu$ discrepancy has been confirmed by the lattice calculation for  the hadronic light-by-light scattering contribution~\cite{Blum:2019ugy}. The measurement of the radiative corrections to the pion form factor also confirms the requirement of a beyond SM contribution to $a_\mu$~\cite{Campanario:2019mjh}.

The situation with the anomalous magnetic moment of the electron, which agreed with its SM prediction at $1.7\sigma$ until recently, is also intriguing. A recent improvement in the measurement of the fine structure constant $\alpha$,\footnote{A very recent measurement \cite{Morel2020} of $\alpha$ contradicts this measurement at $\sim5.4\sigma$ as well as their own past measurement at $\sim 2.5\sigma$. This predicts a $\delta a_e$ at $+1.6\sigma$ as opposed to $-2.5\sigma$ deviation from SM. While we await the resolution of this apparent tension \cite{Muller2020}, positive as well as negative values of $\delta a_e$ at any magnitude are very easily accommodated in the proposed model.} \cite{Parker:2018vye} changed the SM prediction of $(g-2)_e$, which puts its experimentally measured value \cite{Hanneke:2008tm} at a $\sim 2.5\sigma$  deviation
\begin{equation}
\delta a_e = a_e^{\rm exp} - a_e^{\rm SM} = - 8.8(3.6)\times 10^{-13}.
\end{equation}
The most striking feature of these results is the fact that the deviation in muon and electron $(g-2)$ come with opposite signs and also violate the natural scaling of $\sim m_{\mu}^2/m_e^2$. This flies in the face of the Minimal Flavor-violation (MFV) hypothesis and is a clear indication of lepton-flavor non-universality. 

The issue of the muon and electron anomalous magnetic moment have been addressed with an additional $Z^\prime$ gauge boson\cite{Baek:2001kca,Ma:2001md,Heeck:2011wj,Crivellin:2018qmi,Correia:2019woz, CarcamoHernandez:2019ydc,Chen:2020jvl,Haba:2020gkr,Hati:2020fzp,Abdullah:2019ofw,Chao:2020qpe}. However, the $Z^\prime$ contribution as the singular source of the lepton $(g-2)$ is always of the same sign unless the nature of the gauge boson-lepton couplings switch over generations\cite{Jegerlehner:2009ry}. In tandem, there have also been ample attempts to resolve the matter of electron and muon magnetic moments within a supersymmetric framework\cite{Moroi:1995yh,Chattopadhyay:2000ws,Martin:2001st,Hiller:2010ib,Giudice:2012ms,Chakraborty:2015bsk,Kobakhidze:2016mdx,Borzumati:1999sp,Crivellin:2010ty,Endo:2019bcj,Badziak:2019gaf,Yang:2020bmh,Dutta:2018fge}. Considering that the lepton $(g-2)$ from minimal SUSY sources are also always of the same sign, these proposals utilize the effects of large flavor non-universal A-terms\cite{Borzumati:1999sp,Crivellin:2010ty,Endo:2019bcj,Badziak:2019gaf,Yang:2020bmh} or flavor-violating SUSY-breaking terms\cite{Hiller:2010ib,Dutta:2018fge}. Notwithstanding the fact that such effects could be challenging to incorporate in a consistent SUSY-breaking mechanism, the former suffer from the possibilities of large fine-tuning in the lepton mass matrix and charge-breaking minima of the scalar potential. The latter are constrained by strong bounds on charged lepton flavor-violating processes, especially $\tau \rightarrow \mu\gamma$ and $\tau \rightarrow e\gamma$. It has been shown to be possible to explain both the anomalies even without explicit charged lepton flavor violation in a 2HDM model with vector-like fermions~\cite{Chun:2020uzw}, or with an enlarged flavor structure\cite{Hiller:2019mou}. However, the latter scenario requires considerably large yukawa couplings possibly invoking non-perturbative regime.

Inspired by this tantalizing matter of the leptonic MDMs and the constraints on the proposed models, we show that gauged \newu~\cite{He:1990pn,He:1991qd} models embedded within a supersymmetric framework can satisfactorily explain both the electron and muon anomalous magnetic moments simultaneously. And it does so without giving rise to any charged lepton flavor-violating effects, with flavor-universal trilinear soft-terms. Neither does it incorporate any new interactions that could be in conflict with any stability considerations or may have been probed by any present experiments.

We propose a very simple scenario where the sleptons have a certain hierarchy between the generations driven by the corresponding soft SUSY-breaking masses, not unlike the one we see among the leptons themselves. As a result, the $\tilde{\mu}$ and the $\sneumu$ are heavier than the selectron and the $\sneue$ and hence the SUSY contribution to muon $(g-2)$ is suppressed compared to that for the electron. On the other hand, similar contribution to the $(g-2)_e$ is quite sizeable owing to the light $\tilde{e}$ and $\sneue$ masses. The muon $(g-2)$ has a different source, viz. the $L_\mu-L_\tau$ gauge boson $Z^\prime$. This interplay allows us to explain both the muon and electron anomalous magnetic moment measurements, otherwise impossible in either \newu or SUSY models by themselves. We find a viable parameter space satisfying the bounds on the masses of Sparticles from the LHC \cite{Tanabashi:2018oca}. These bounds and the criteria for satisfying both the lepton $(g-2)$ places an upper bound on the mass of the selectron, subject to the choice of the $\mu$-parameter and the ratio of Higgs VEVs, as well as a lower bound on the smuon mass. Most importantly, the scenario can be probed most certainly in the near future by the ongoing experiments at the LHC. 

In addition, the low mass $Z^\prime$ which explains $(g-2)_\mu$ are probed at the ongoing coherent elastic neutrino-nucleus scattering experiments (CE$\nu$NS), e.g., COHERENT and Coherent CAPTAIN-Mills (CCM). The recent data from the LAr~\cite{Akimov:2020pdx} and CsI~\cite{Akimov:2017ade} detectors at the COHERENT experiment provide important constraints alongside the existing constraints emerging from CCFR~\cite{Mishra:1991bv}, Borexino~\cite{Bellini:2013lnn}, BaBaR~\cite{TheBABAR:2016rlg}, supernova~\cite{Croon:2020lrf} etc. on the $Z^\prime$ mass and coupling. The ongoing CCM experiment with its large fiducial mass liquid Argon detector is projected to register a significantly large number of CE$\nu$NS events per year\cite{ccm1,ccm2} and hence be able to probe much of the pertinent parameter space.

We describe the model in Section \ref{sec:model} and discuss the additional sources of the lepton magnetic moments in it in Section \ref{sec:madm}. Section \ref{sec:const} discusses the constraints from LHC that are pertinent to the parameter space where both lepton magnetic moments can be explained. This section also describes the parameter ranges over which the numerical scans were conducted as well as the motivations for the choices. Our results are summarized in Section \ref{sec:result} alongside the figures from our numerical scans and finally, we conclude in Section \ref{sec:conclusion}.

\section{The model}\label{sec:model}

A supersymmetric gauged \newu model requires at least two additional $R$-Parity even superfields $\hat{\eta}$ and $\hat{\etabar}$ to be consistent with $Z$-boson decay observations and neutrino mixing texture \cite{Banerjee:2018eaf}. Scalars corresponding to these \newu charged additional superfields, singlets under SM gauge symmetries and $R$-Parity even, acquire vacuum expectation values (VEVs) to break the additional gauge symmetry spontaneously. The cancellation of chiral anomalies require these superfields to always have equal and opposite charge under the \newu symmetry. While $R$-Parity was not considered to be a global symmetry in \cite{Banerjee:2018eaf}, we work in a simplified scenario where $R$-Parity is conserved and the superpotential is given by
\begin{equation}
W= W_{\rm MSSM} -\mu_\eta \hat{\eta} \hat{\etabar}
\end{equation}
where $W_{\rm MSSM}$ is the MSSM superpotential with flavor-diagonal yukawa couplings in the leptonic sector.
The charge assignments for the chiral superfields charged under \newu are given in Table \ref{table:model-charges}, the unlisted superfields are singlets under the additional gauged symmetry. Notably, yukawa terms like $\hat{\eta}\hat{L}_{\mu}\hat{H}_U$ and $\hat{\etabar}\hat{L}_{\tau}\hat{H}_U$, although gauge singlets, break $R$-Parity and hence are forbidden. Their presence by means of alternate $R$-Parity assignments for the $\hat{\eta}/\hat{\etabar}$ superfields would lead to spontaneous breaking of $R$-Parity alongside the \newu symmetry.

\begin{table}[t]
\centering
\begin{tabular}{|c|cccc|cc|}
\hline\hline
Superfields & ${\hat L}_\mu$ & ${\hat E}^c_\mu$ & ${\hat L}_\tau$ & ${\hat E}^c_\tau$ & $\hat\eta$ & $\hat {\bar \eta}$ \\
\hline
$U(1)_{L_\mu - L_\tau}$ & 1 & -1 & -1 & 1 & -1 & 1 \\
\hline\hline
\end{tabular}
\caption{${L_\mu - L_\tau}$ charge assignments for the chiral superfields that are not singlets under the additional symmetry.}
\label{table:model-charges}
\end{table}

The soft SUSY-breaking terms for our specific choice of parameters is given by
%\begin{widetext}
\begin{eqnarray}\label{e:mini:soft}
-\mathcal{L}_{\rm soft}&=&\left(\frac{M_3}{2}(i\gino)(i\gino) + \frac{M_2}{2}(i\wino)(i\wino) +
\frac{M_1}{2}(i\bino)(i\bino) + \frac{M_0}{2}(i\bpino)(i\bpino) + h.c\right) \nn
&-& M_{10}(i\bino)(i\bpino)+A \bigg(y_u^{ij} H_u \sQ_j\sU_i - y_d^{ij} H_d \sQ_j\sD_i  - y_l^i H_d\sL_i\sE_i+ h.c\bigg)\nn
&+& M_Q^2 \bigg(\sQ^{\dagger}\sQ + \sU^{c\dagger}\sU^c + \sD^{c\dagger}\sD^c \bigg)+ \sum_{l = e, \mu, \tau} \big(M_{\tilde{l}_L}^2 \sL_l^{\dagger}\sL_l + M_{\tilde{l}_R}^2 \sE_l^{c\dag}\sE_l^c \big)\nn 
&+& M^2_{H_d} H_d^{\dagger} H_d + M^2_{H_u} H_u^{\dagger} H_u + M_{\eta}^2 \eta^\dagger \eta + M_{\etabar}^2 \etabar^\dagger \etabar  -\left(B_0 H_d H_u + B_\eta \eta \etabar  + H.c.\right).
\end{eqnarray}
%\end{widetext}

Conservation of $R$-Parity ensures that there is no generation-mixing between the sleptons, nor do they mix with the other scalars in the spectrum. Apart from the additional gauge boson and its fermion superpartner, the fermions and bosons from the two additional superfields augment the MSSM neutralino, scalar and pseudoscalar spectrum. The soft SUSY-breaking masses corresponding to the two Higgses ($H_d$ and $H_u$) and the additional fields ($\eta$ and $\etabar$) are not independent parameters but are solved for by minimizing the scalar potential. We list here the most pertinent mass matrices and the rotation relations from mass to gauge eigenstate basis.

The neutralino mass matrix written in the basis: \( \left(i\bino, i\wino, \tilde{H}_d^0, \tilde{H}_u^0, i\bpino, \tilde{\eta}, \tilde{\etabar}\right) \) is,
%\begin{widetext} 
\begin{equation} 
m_{\tilde{\chi}^0} = \left( 
\begin{array}{ccccccc}
M_1 &0 &-\frac{1}{2} g_1 v_d  &\frac{1}{2} g_1 v_u  &M_{10} &0  &0 \\ 
0 &M_2 &\frac{1}{2} g_2 v_d  &-\frac{1}{2} g_2 v_u  &0 &0 &0\\ 
-\frac{1}{2} g_1 v_d  &\frac{1}{2} g_2 v_d  &0 &- \mu  &-\frac{1}{2} g_{Y X} v_d  &0 &0\\ 
\frac{1}{2} g_1 v_u  &-\frac{1}{2} g_2 v_u  &- \mu  &0 &\frac{1}{2} g_{Y X} v_u  &0 &0\\ 
M_{10} &0 &-\frac{1}{2} g_{Y X} v_d  &\frac{1}{2} g_{Y X} v_u  &M_{0} &- g_{X} v_{\eta}  &g_{X} v_{\bar{\eta}} \\ 
0 &0 &0 &0 &- g_{X} v_{\eta}  &0 &\mu_{\eta}\\ 
0 &0 &0 &0 &g_{X} v_{\bar{\eta}}  &\mu_{\eta} &0\end{array} 
\right) 
 \end{equation} 
This matrix is diagonalized by the unitary matrix \(N\): 
\begin{equation} 
N^* m_{\tilde{\chi}^0} N^{\dagger} = m^{dia}_{\tilde{\chi}^0} 
\end{equation} 
with 
\begin{align} 
&{\tilde{B}} = \sum_{j}N^*_{j 1}\tilde{\chi}^0_{{j}}\,, \hspace{1cm} 
\tilde{W}^0 = \sum_{j}N^*_{j 2}\tilde{\chi}^0_{{j}}\,, \hspace{1cm} 
\tilde{H}_d^0 = \sum_{j}N^*_{j 3}\tilde{\chi}^0_{{j}}\,, \hspace{1cm}
\tilde{H}_u^0 = \sum_{j}N^*_{j 4}\tilde{\chi}^0_{{j}}\nn  
&{\tilde{B}^\prime} = \sum_{j}N^*_{j 5}\tilde{\chi}^0_{{j}}\,, \hspace{1cm} 
\tilde{\eta} = \sum_{j}N^*_{j 6}\tilde{\chi}^0_{{j}}\,, \hspace{1cm}  
\tilde{\etabar} = \sum_{j}N^*_{j 7}\tilde{\chi}^0_{{j}}.
\end{align}
%\end{widetext}
The chargino mass matrix is exactly same as the one for MSSM. Written in the basis: \( \left(\tilde{W}^-, \tilde{H}_d^-\right), \left(\tilde{W}^+, \tilde{H}_u^+\right) \), the mass matrix $m_{\tilde{\chi}^\pm}$ is diagonalized by the unitary matrices $U$ and $V$ such that, 
\begin{equation} 
U^* m_{\tilde{\chi}^\pm} V^{\dagger} = m^{dia}_{\tilde{\chi}^\pm} 
\end{equation} 
with 
\begin{align} 
&\tilde{W}^- = \sum_{j}U^*_{j 1}\tilde{\chi}^-_{{j}}\,, \hspace{1cm} 
\tilde{H}_d^- = \sum_{j}U^*_{j 2}\tilde{\chi}^-_{{j}}\nn 
&\tilde{W}^+ = \sum_{j}V^*_{1 j}\tilde{\chi}^+_{{j}}\,, \hspace{1cm} 
\tilde{H}_u^+ = \sum_{j}V^*_{2 j}\tilde{\chi}^+_{{j}}.
\end{align}

The slepton mass-squared matrices are given by,
\begin{equation} 
m^2_{\tilde{l}} = \left( 
\begin{array}{cc}
m^2_{\tilde{l}_L\tilde{l}_L^*} &m_l\big(A-\mu \tan\beta\big)\\ 
m_l\big(A-\mu \tan\beta\big) &m^2_{\tilde{l}_R\tilde{l}_R^*}\end{array} 
\right) 
 \end{equation} 
where 
\begin{eqnarray}
m^2_{\tilde{l}_L\tilde{l}_L^\ast} &=& M^2_{\tilde{l}_L} + Q_X^l M_{Z^\prime}^2 \cos\left(2\gamma\right)+ \left(\frac12 -c^2_W\right)M^2_Z \cos\left(2\beta\right),\nn
m^2_{\tilde{l}_R\tilde{l}_R^\ast} &=& M^2_{\tilde{l}_R} + Q_X^l M_{Z^\prime}^2 \cos\left(2\gamma\right)- \left(1 -c^2_W\right)M^2_Z \cos\left(2\beta\right).
\end{eqnarray}
Here $\tan\gamma = {\langle \eta \rangle}/{\langle \etabar \rangle}$ and $\tan\beta = {\langle h_u \rangle}/{\langle h_d \rangle}$ are the ratio of the corresponding scalar VEVs. $Q_X^l$ are the charges of the respective leptons under the gauged \newu symmetry.

This matrix is diagonalized by unitary matrices \(X^l\): 
\begin{equation} 
X^l m^2_{\tilde{l}} X^{l\dagger} = m^{2}_{\tilde{l}, dia} 
\end{equation} 
with 
\begin{align} 
\tilde{l}_L = \sum_{j}X^{l\ast}_{j 1}\tilde{l}_{{j}}\,, \hspace{1cm} 
\tilde{l}_R = \sum_{j}X^{l\ast}_{j 2}\tilde{l}_{{j}}.
\end{align} 
The sneutrino mass matrix is diagonal in the gauge eigenstate basis itself.

While most of the parameters were left free, the sign of the $\mu$-term is kept negative to get a negative SUSY contribution to the lepton magnetic moments. The squark and gluino masses have been set to be decouplingly large and the stau masses are kept $\sim 17.8$ times the smuon masses\footnote{Changing this ratio does not affect our analysis or our results as long as m$_{\tilde{\tau}}$ $\gtrsim$ m$_{\tilde{\mu}}$. Its effect on the $L_\mu - L_\tau$ parameter space is discussed in Section \ref{sec:madm}}, replicating the corresponding lepton hierarchy. We assume zero tree level kinetic mixing ($g_{Y X}$), however it is unavoidably generated radiatively via loops involving the second and third generation leptons and sleptons.

\section{Anomalous Magnetic Moment}\label{sec:madm}

\begin{figure}[t]
%\centering
\subfigure[\label{fig:mdmdiagcharg}]{\includegraphics[width=4.7cm]{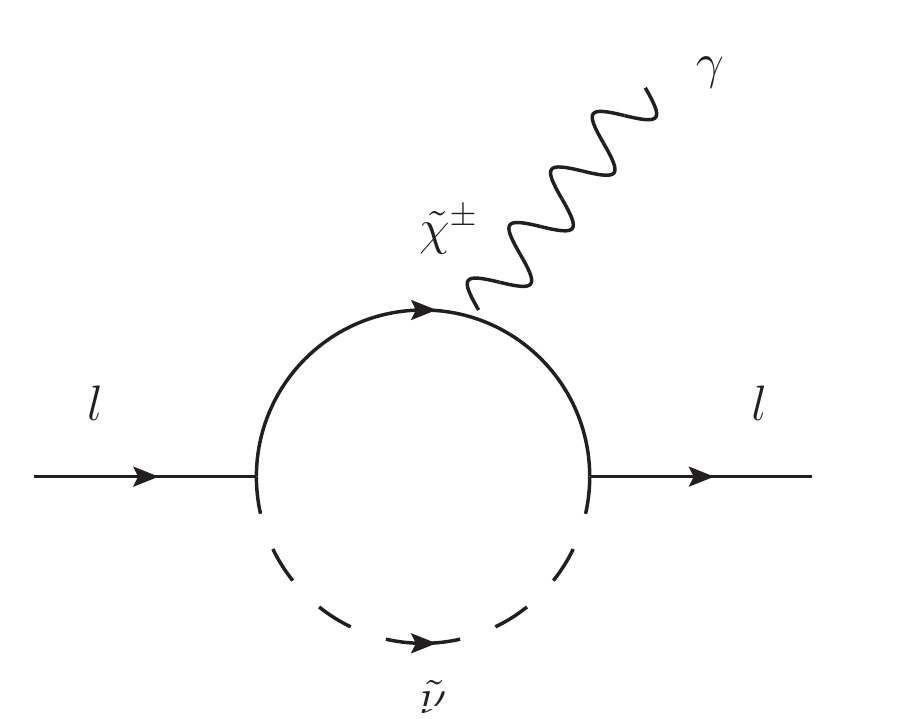}}
\subfigure[\label{fig:mdmdiagneut}]{\includegraphics[width=4.7cm]{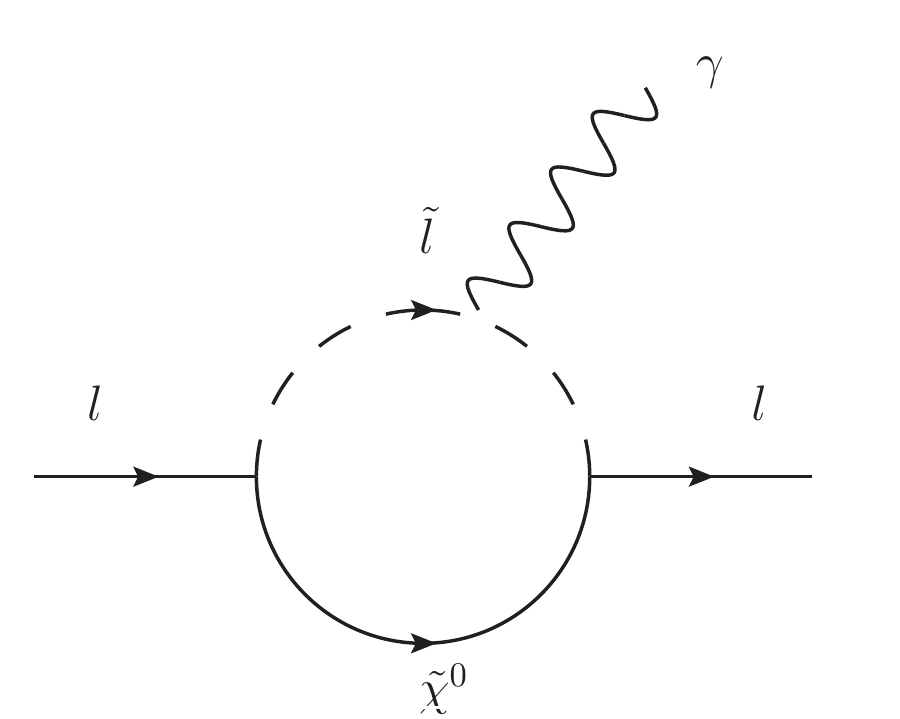}}
\centering
\subfigure[\label{fig:mdmdiagzprime}]{\includegraphics[width=4.7cm]{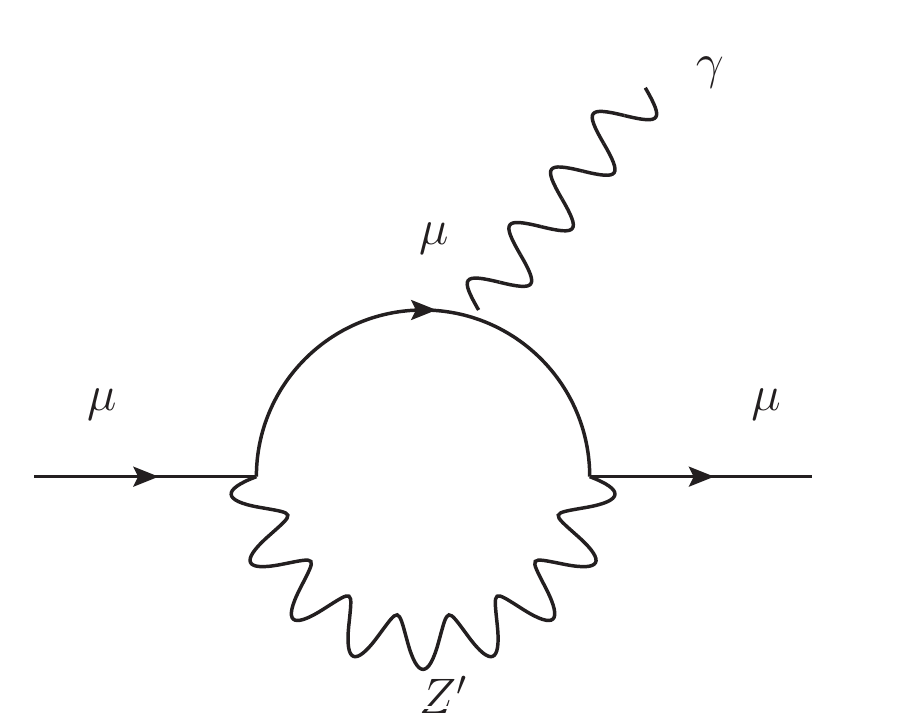}}
\caption{Figures \ref{fig:mdmdiagcharg} and \ref{fig:mdmdiagneut} constitute the SUSY contribution to general lepton magnetic moments. Figure \ref{fig:mdmdiagzprime} contributes solely to the anomalous magnetic moment of the muon\label{fig:mdmdiags}}
\end{figure}

Supersymmetry unavoidably tacks on additional contributions to the SM corrections to the lepton magnetic moments. Even the Minimal Supersymmetric Standard Model (MSSM) adds contributions to $g-2$ that have been shown to be enough to explain the discrepancy of the $(g-2)_{\mu}$ from its SM prediction. Although the lack of vindication from various experimental SUSY searches constricts the domain of validity of this statement over time, the muon $g-2$ is an ideal observable to constrain some parameters of the model. For instance, the sign of the ``$\mu$-term", the ratio of the Higgs VEVs ($\tan\beta$)and the mass scale of the sparticles crucially determine the sign and magnitude, respectively, of the lepton $(g-2)$.

The two main components of the SUSY contribution to the lepton $(g-2)$ are the slepton-neutralino and the chargino-sneutrino loops shown in Figs.\ref{fig:mdmdiagcharg} and \ref{fig:mdmdiagneut}. When the mass scale of the 
superpartners are roughly of the scale $M_{SUSY}$, the total contribution is given in the MSSM by \cite{Martin:2001st,Moroi:1995yh},
\begin{equation}
\Delta a_{l}^{MSSM}=14\frac{m_l^2}{m_{\mu}^2}\text{ Sign}(\mu)\tan \beta\left(\frac{100 \text{ GeV}}{M_{SUSY}}\right)^2 10^{-10}.
\end{equation}
Note that the dominant supersymmetric lepton $(g-2)$ contributions are from the chargino-sneutrino loop. Additionally, they are of the same sign and scale exactly by the mass squared of the corresponding lepton, flying in the face of the present experimental observations.

Using the mixing matrices in section \ref{sec:model}, the neutralino-slepton loop contribution to the lepton ($g-2$) was found to be\cite{Martin:2001st,Moroi:1995yh},
\begin{eqnarray}
a_{l}^{\chi^0}=&&-\frac{m_{l}}{16\pi^2}\sum_{i=1}^{7}\sum_{j=1}^{2}\bigg[\big(|n^L_{ij}|^2+|n^R_{ij}|^2\big)\frac{m_{l}}{12{m_{\tilde{l}}^2}_j}F_1^N(x_{ij})\nonumber\\
&&+\frac{{m_{\tilde{\chi}^0}}_i}{3{m_{\tilde{l}}^2}_j}\text{Real}\big(n^L_{ij}n^R_{ij}\big)F_2^N(x_{ij})\bigg]
\end{eqnarray}
where,
\begin{eqnarray}
F_1^N(x)&=&\frac{2(1-6x+3x^2+2x^3-6x^2\log x)}{(1-x)^4}\nonumber\\
F_2^N(x)&=&\frac{3(1-x^2+2x\log x)}{(1-x)^3}
\end{eqnarray}
with
\begin{equation*}
x^l_{ij}=\frac{{m_{\tilde{\chi}^0}}_i^2}{{m_{\tilde{l}}^2}_j}
\end{equation*}
and 
\begin{eqnarray}
n^L_{ij}=&&-y_{l} N^{\ast}_{i,3}X^{l\ast}_{j,1} - {\surd 2} g_1 N^{\ast}_{i,1}X^{l\ast}_{j,2} + {\surd 2} g_X N^{\ast}_{i,5}X^{l\ast}_{j,2}\nn
n^R_{ij}=&&-y_{l} N_{i,3} X^{l\ast}_{j,2} + \big(\frac{g_1}{\surd 2} N_{i,1}+ \frac{g_2}{\surd 2} N_{i,2}- {\surd 2} g_X N_{i,5}\big)X^{l\ast}_{j,1}.\nonumber
\end{eqnarray}

The contribution of the chargino-sneutrino loop to the lepton ($g-2$) is\cite{Martin:2001st,Moroi:1995yh},
\begin{eqnarray}
a_{l}^{{\tilde{\chi}}^{\pm}}=&&\frac{m_{l}}{16\pi^2}\sum_{k=1}^2\bigg[\frac{m_{l}}{12m_{\tilde{\nu}_l}^2}\big(|c^{L}_{k}|^2
+|c^{R}_{k}|^2\big)F_1^C(y^l_{k}) \nn
&& +\frac{2m_{{\tilde{\chi}}_k^{\pm}}}{3m_{\tilde{\nu}_l}^2}\text{Real}\big(c^{L}_{k}c^{R}_{k}\big)F_2^C(y^l_{k})\bigg]
\end{eqnarray}
where,
\begin{eqnarray}
F_1^C(x)=&&\frac{2(2+3x-6x^2+x^3+6x\log x)}{(1-x)^4}\nn
F_2^C(x)=&&-\frac{3(3-4x+x^2+2\log x)}{2(1-x)^3}
\end{eqnarray}
with
\begin{equation*}
y^l_{k}=\frac{m_{{\tilde{\chi}}_k^{\pm}}^2}{m_{\tilde{\nu}_l}^2}
\end{equation*}
and
\begin{eqnarray}
c^{L}_{k}=y_{l}U^{\ast}_{k2}, \qquad c^{R}_{k}=-g_2 V_{k1}.\nonumber
\end{eqnarray}

Apart from the supersymmetric contribution to $\Delta a_{l}$, the $Z^{\prime}$ boson adds an essential part to the muon magnetic moment, as shown in Fig.\ref{fig:mdmdiagzprime}. It is given by\cite{Baek:2001kca,Ma:2001md,Heeck:2011wj}
\begin{equation}
\Delta a_{\mu}^{Z^{\prime}} = \frac{g_X^2m_{\mu}^2}{4\pi^2}\int_0^1 dz \frac{z^2(1-z)}{m_{\mu}^2z+M_{Z^{\prime}}^2(1-z)}.  
\end{equation}
This contribution to $(g-2)_{\mu}$ is always positive. 

\begin{figure}[t]
\centering
\includegraphics[scale=0.55]{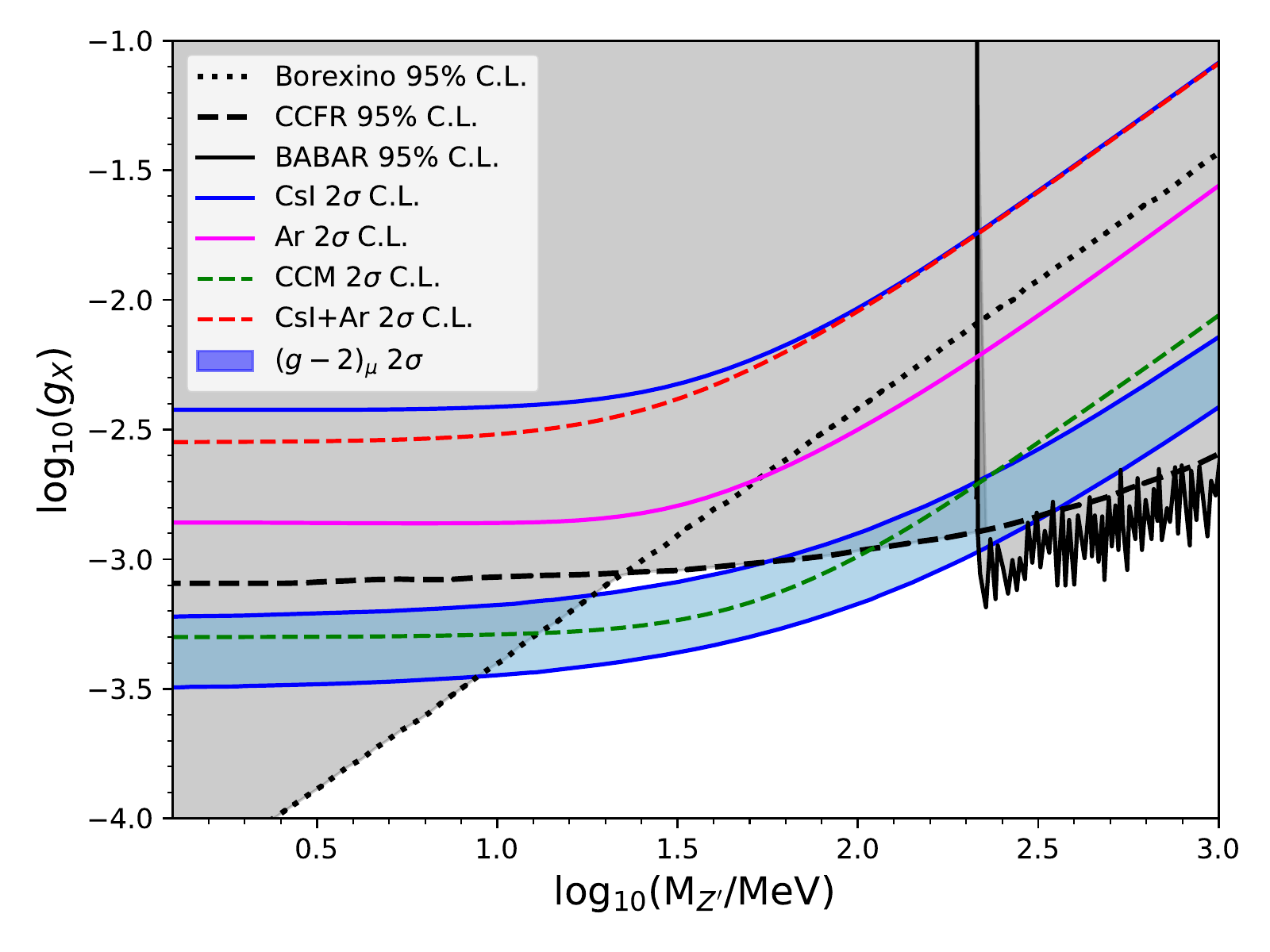}
\caption{The parameter space of interest in \newu models where $(g-2)_\mu$ can be explained by the $Z^\prime$ contribution alone is highlighted in blue. Strongest present constraints\cite{Mishra:1991bv,Bellini:2013lnn,TheBABAR:2016rlg,Sirunyan:2018nnz} are shown alongside the highlighted region. The COHERENT CsI, Ar and combined limits are from a likelihood analysis\cite{Banerjee:2021laz} of the recently released full data. The Coherent CAPTAIN-Mills (CCM) exclusion is shown assuming 3 years of data taking at the running experiment.\label{fig:mzpgx}}
\end{figure}

In our analysis we have kept the sign of the $\mu$-term to be negative, resulting in a negative SUSY contribution to $\delta a_l$. In order for this to be able to explain the $(g-2)_e$ ($\sim 10^{-12}$), we require the first generation sleptons to be as light as a few hundred GeV. If this were the general mass scale of all the sleptons, notwithstanding the collider constraints, the negative SUSY contribution to the $(g-2)_{\mu}$ would totally swamp the positive $Z^\prime$ contribution. It is imperative then that to satisfy both lepton anomalous magnetic moments, we require the first generation sleptons to be sufficiently light and, at the same time, the second generation sleptons be heavy enough. This means that the region of $L_\mu-L_\tau$ parameter space satisfying $(g-2)_{\mu}$ in this case is the same as that for non-SUSY \newu models (see Fig.\ref{fig:mzpgx}). Yet, it is now possible to satisfy $(g-2)_e$ as well within that region. For a recent study of the present constraints on this class of models and the parameter space yet to be explored by proposed future experiments see, for example, Ref. \cite{Amaral:2020tga}. 

The COHERENT collaboration recently published the full set of data corresponding to their observation of coherent neutrino nucleus scattering (CE$\nu$NS) on both CsI and Ar detectors\cite{Akimov:2018vzs,Akimov:2020pdx,Akimov:2020czh}. A rigorous likelihood analysis\cite{Banerjee:2021laz} of the data including energy, timing as well as the pulse shape discriminator (PSD) data led to the exclusions shown in Fig.\ref{fig:mzpgx}. The Collaboration has recently announced the imminent release of additional data from its CsI detector. The additional data finds a best fit CE$\nu$NS count of 306 events against a SM prediction of 333. While the preference for reductive BSM effect persists in this updated dataset, the SM deviation seems to be smaller. This should lead to the combined exclusions getting stronger, however, whether it will be able to probe the unconstrained region requires an analysis with the awaited full data. A similar likelihood based approach was used to show projected exclusions from the CCM experiment that is presently running at the Los Alamos National Laboratory~\cite{ccm1,ccm2}.
The experiment uses neutrinos produced by an 800 MeV proton beam at 20 Hz ($\sim$ 80 kW) impinged on a tungsten target to detect CE$\nu$NS with a liquid Argon detector of large fiducial mass (7 tons) kept at 20m. We assume 5000 hours of operation per year, equivalent to $\sim 3\times 10^{22}$ protons-on-target with 0.0425 neutrinos per proton for each flavor, for three years and a nuclear recoil energy detector window of 25 - 150 KeV. Although the exclusions here are shown for $m_{\tilde{\tau}}/m_{\tilde{\mu}} = m_\tau/m_\mu$, this ratio has only a very mild effect on them since the kinetic mixing parameter has a logarithmic dependence on it\cite{Banerjee:2018mnw}. Larger values of this ratio excludes a larger region of the parameter space. Changes in the kinetic mixing parameter arising due to changes in this ratio could, in general, affect the exclusions from Borexino as well but this dependence is also mild.  

\section{Constraints from LHC and Numerics}\label{sec:const}

\begin{figure}[t]
%\centering
\subfigure[\label{fig:const1}]{\includegraphics[width=4.8cm]{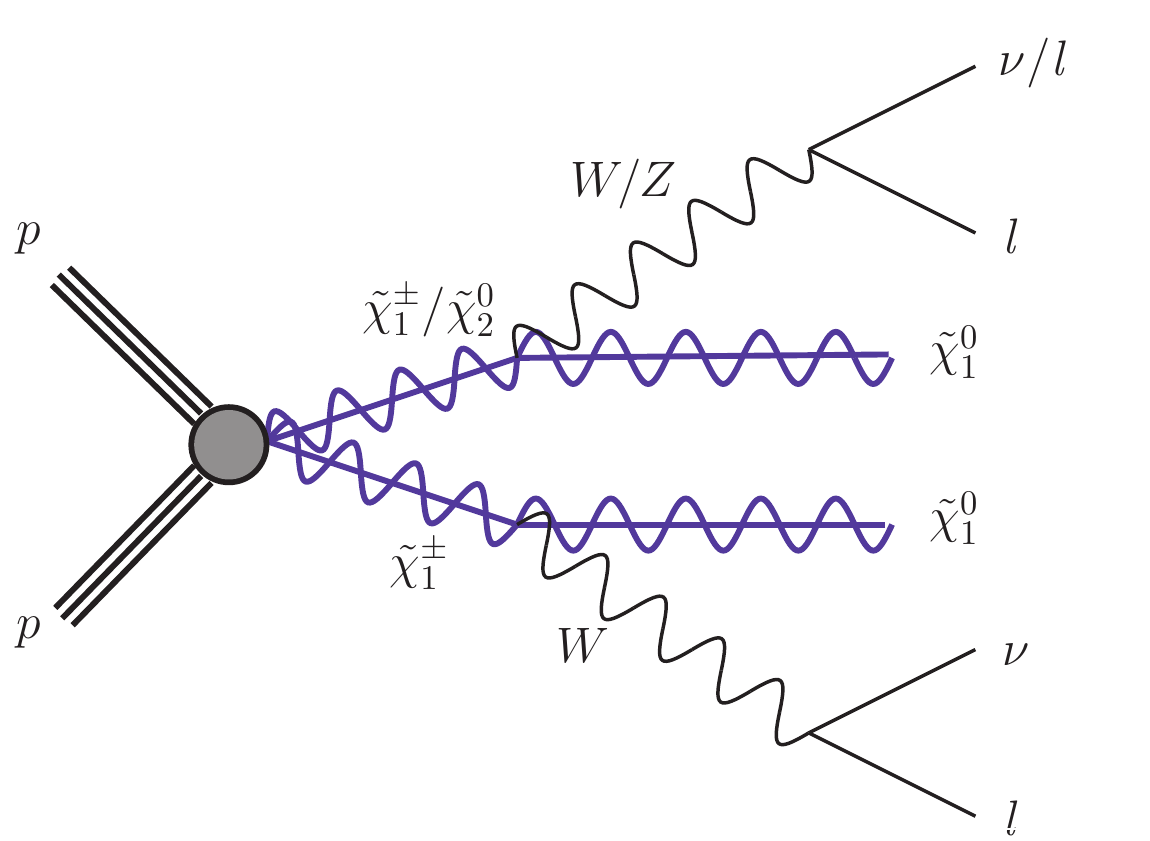}}
\subfigure[\label{fig:const2}]{\includegraphics[width=4.8cm]{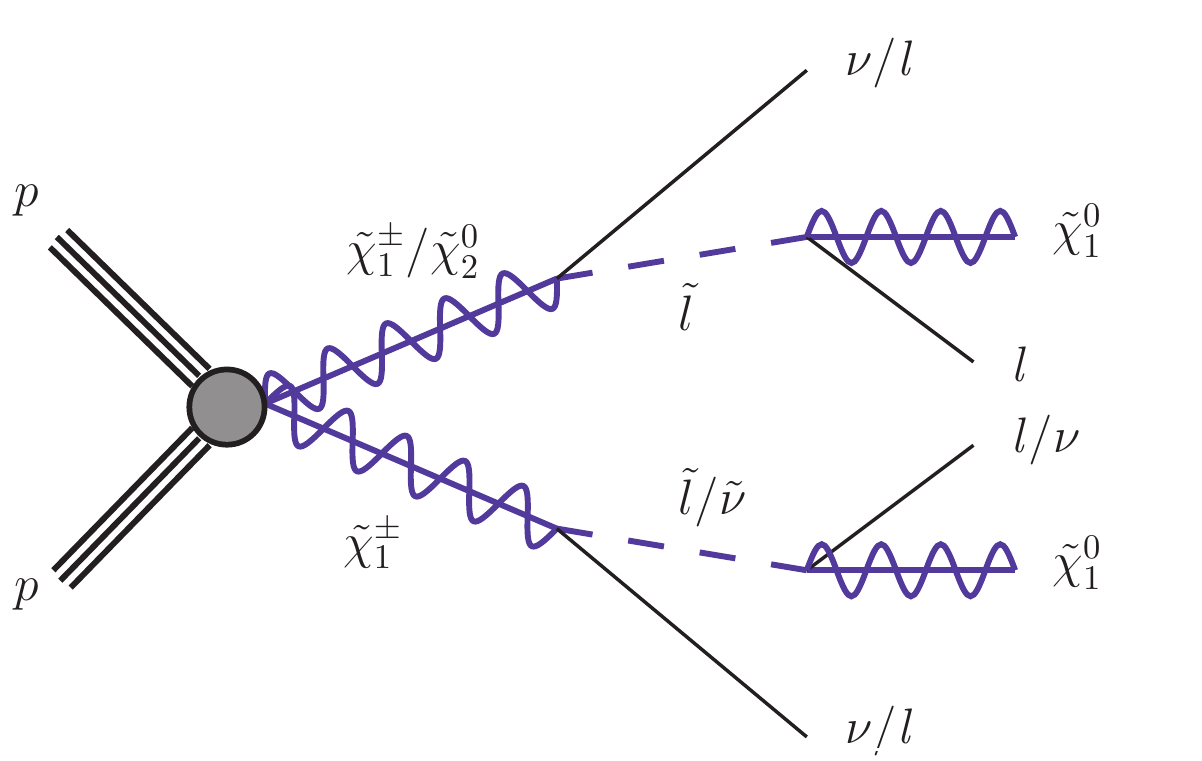}}
\centering
\subfigure[\label{fig:const3}]{\includegraphics[width=4.8cm]{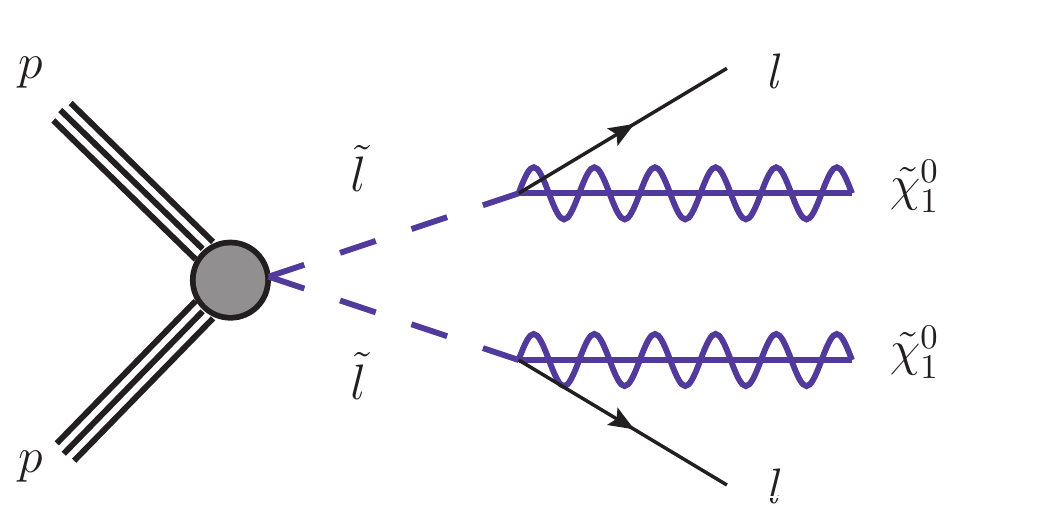}}
\caption{Diagrams corresponding to the SUSY searches at the ATLAS detector, with two and three leptons in the final state alongside missing energy, which put the strongest constraints on the model: \ref{fig:const1} $W/Z$ mediated decays, \ref{fig:const2} $\tilde{l}$/$\tilde{\nu}$ mediated decays and \ref{fig:const3} direct slepton pair production. Ref.\cite{Aaboud:2018jiw} includes jets from $W$-decays in their analysis alongside fig.\ref{fig:const1} when considering the pair production of $\tilde{\chi}^\pm_1$ and $\tilde{\chi}^0_2$. Ref.\cite{Aad:2019vnb} includes intermediate $\tilde{\nu}$'s in their analysis of $\tilde{\chi}^\pm_1$ pair production.\label{fig:consts}. We also use bounds from $W/h$ mediated decays of a chargino/neutralino pair\cite{Aad:2020qnn} that proceeds via the same diagram as fig.\ref{fig:const1} with one of the gauge boson fields replaced by the higgs boson. Diphoton and $b\bar{b}$ decay modes of the higgs boson have been considered.}
\end{figure}

The requirement for the first generation sleptons to be considerably light in order to explain the $(g-2)_e$ imposes strong constraints on the model from the LHC $2l + 3l$ final state searches. We discuss the nature of these constraints and the parameters of the numerical scans used to find a viable scenario in this section.

The strongest bounds on the lightest neutralino-lighter chargino mass-plane come from the searches for the electroweak production of charginos and neutralinos decaying into two lepton and three lepton final states (see Figs.\ref{fig:const1} and \ref{fig:const2}) at the ATLAS detector\cite{Aaboud:2018jiw,Aad:2019vnb}. There are two different situations to consider in this case, and depending upon the scenario in question, the constraints change. The most severe constraints come from the combined $2l + 3l$ searches assuming $\tilde{\chi}^0\tilde{\chi}^\pm$ pair production and their subsequent decay via $\tilde{l}/\tilde{\nu}$. The lower bounds on the $\tilde{\chi}^0$ and $\tilde{\chi}^\pm$ masses in this case are sometimes as high as 700 GeV and 1.2 TeV respectively. However the constraints are much weaker if one considers the case where the $\tilde{l}/\tilde{\nu}$ is heavier than the lightest $\tilde{\chi}^\pm$. In such situations the same final states occur through $W/Z$ or $W/h$ mediated decays(see Fig.\ref{fig:const1}) \cite{Aaboud:2018sua,Aad:2020qnn}. The bounds for scenarios involving compressed spectra need to be adapted separately\cite{Aad:2019qnd}.

Similarly, the strongest bounds on the lightest neutralino-slepton mass-plane come from the searches of direct slepton pair production and their subsequent decay into two lepton final states as shown in Fig.\ref{fig:const3} from the same study\cite{Aaboud:2018jiw,Aad:2019vnb,Aad:2019qnd}. We use single generation bounds on combined left and right-handed sleptons in order to constrain selectron and smuon masses. This distinction is crucial in our study as the premise of explaining both lepton MDM's require light selectrons and heavy smuons. Constraints on chargino, neutralino and slepton masses from similar searches at the CMS detector \cite{Sirunyan:2017qaj, Sirunyan:2018lul, Sirunyan:2019iwo, Sirunyan:2020eab} are either weaker or of a similar nature. Additionally, we incorporate bounds on the lightest chargino mass and lifetime from long-lived charged particle searches\cite{Aaboud:2017mpt, Aad:2015qfa, Aad:2013yna}.

The exclusions on slepton masses assume mass degenerate sleptons and their MSSM decay modes only. In our model however, the decays of the second and third generation sleptons may be modified through their couplings with the additional gauge boson. The decay width of the slepton due to effects beyond the MSSM in our model is given by\cite{Haber:1984rc},  
\begin{eqnarray}
\Gamma_{\rm NP}[\tilde{l}\rightarrow {\rm l}\tilde{\chi}^0]&=&g_X^2 \vert N_{15}\vert^2 \frac{p_f}{8\pi M_{\tilde{l}}^2}\bigg[\left(\sum_{i,j=1}^2\vert X^l_{ij}\vert^2\right)\left(M_{\tilde{l}}^2-m_l^2-M_{\tilde{\chi}^0}^2\right)\nn
&-& 4 m_l M_{\tilde{\chi}^0} {\rm Re}[\sum_{i=1}^2 X^{l\ast}_{i1}X^l_{i2}]\bigg].
\end{eqnarray}
$p_f$ being the magnitude of the three-momentum of the decay products in the rest frame of the slepton and $N_{15}$ is the $\tilde{B}^\prime$ mixing element in the lightest neutralino. This implies that,
\begin{equation}
\frac{\Gamma_{\rm NP}}{\Gamma_{\rm MSSM}}\sim \frac{g_X^2}{4\pi\alpha} \lesssim 2.7\times 10^{-6},
\end{equation}
that is, the modification to the branching ratio of the sleptons due to the additional gauge boson is all but negligible.

Although the $\tilde{\eta}/\tilde{\etabar}$ dominated neutralinos are decoupled, the $\bpino$ mass is left unrestricted. This may, in principle, modify the decay chains assumed in the above exclusions. However, the couplings between the MSSM-like fields and $\bpino$ is negligible in comparison with their other couplings. Its presence, thus, goes largely unnoticed in both the collider and the $(g-2)$ analysis. Still, when the lightest supersymmetric particle (LSP $\tilde{\chi}^0_1$) is $\bpino$-like, the NLSP($\tilde{\chi}^0_2$), which is then necessarily MSSM-like, can decay radiatively into a photon and the LSP. This decay width is given by\cite{Haber:1988px},
\begin{equation}
    \Gamma(\tilde{\chi}^0_2 \rightarrow \tilde{\chi}^0_1 \gamma) = (g_{\tilde{\chi}^0_2\tilde{\chi}^0_1\gamma})^2\frac{(M_{\tilde{\chi}^0_2}^2 - M_{\tilde{\chi}^0_1}^2)^3}{8\pi M_{\tilde{\chi}^0_2}^5},
\end{equation}
where $g_{\tilde{\chi}^0_2\tilde{\chi}^0_1\gamma}$ is computed by evaluating all the relevant one-loop diagrams.
Depending on the lifetime of the $\tilde{\chi}^0_2$, this decay may affect the Big Bang Nucleosynthesis (BBN) and the Cosmic Microwave Background (CMB). Considering, we exclude scenarios that predict a lifetime greater than $\sim 0.1$ seconds for the NLSP under these circumstances. We also note in conjunction that this lifetime never falls below $\sim 10^{-5}$ seconds. This ensures that the MSSM-like NLSP is stable within collider scales and thus behaves like the LSP when it comes to the LHC while still not modifying the CMB or BBN in any way. The NLSP hence sets the mass scale for the rest of the spectrum as well in this scenario, which in turn results in an upper limit on its mass that comes from the requirement to explain the $(g-2)_e$.

The exclusions mentioned above are all for various simplified scenarios assuming 100$\%$ branching ratios. We recast them with our model specific mixing elements and branching ratios which are all functions of the choice of parameters. The predicted cross section for a specific process is given by,
\begin{equation}
    \sigma^X = \sigma^{\rm G}\times {\rm F}_{\rm M}\times \left(\prod {\rm BR}\right)
\end{equation}
where $\sigma^{\rm G}$ is the pair production cross section for pure gauge eigenstates\cite{Fuks:2012qx,Fuks:2013vua,Bozzi:2007qr,Fuks:2013lya,Fiaschi:2018xdm,Beenakker:1999xh}. This cross section is modified by the mixing element factor ${\rm F}_{\rm M}$ that gives us the production cross section for the physical states. We multiply this with the product of all the branching ratios for a given process to obtain the final predicted cross section $\sigma^X$. This can then be compared with the upper limits\cite{csul1,csul2,csul3,csul4,csul5} on the cross section for the process. We exclude those parameter points that give a predicted cross section larger than the upper limit.

While the constraints on the sleptons and those pertaining to cosmological observations are always applicable, the LHC bounds on the electroweakinos are not all equally pertinent under all circumstances. Considering, we segregate parameter choices into four classes, each of which call for distinct combinations of the described constraints.

\begin{enumerate}
    \item $M_2 < (M_1, \vert\mu\vert)$: Constraints on the long-lived wino-like lightest chargino are applicable. The heavier higgsinos may be produced in pairs. Constraints considering their subsequent decays via SM gauge/Higgs bosons are also applicable.
    \item $M_1 < M_2 < \vert\mu\vert$: Strong bounds on the lightest wino-like chargino and associated neutralino from their pair-production scenarios with decays via either sleptons or SM gauge bosons. Constraints on the heavier higgsinos are negligible.
    \item $M_1 < \vert\mu\vert < M_2 $: Similar constraints as the previous case apart from the fact that constraints on both the lightest higgsino-like and heavier wino-like chargino and their associated neutralinos need to be considered.
    \item $\vert\mu\vert < (M_1, M_2)$: Weak constraints on the lightest nearly degenerate higgsino triplet. Constraints on the heavier wino-like chargino and the associated neutralino are not negligible and are considered.
\end{enumerate}

All the soft SUSY-breaking masses for the gauginos as well as the sleptons and the negative $\mu$-parameter were allowed to vary between 100 GeV-1 TeV in the numerical scan. The scan implements the above constraints and checks if both $(g-2)_\mu$ and $(g-2)_e$ can be explained. $\tan\beta$ was allowed to vary between 10 to 50. The other parameters, like the A and B-parameters were allowed to vary between 500 GeV-1 TeV. The squarks, gluinos and the fermions corresponding to the $\eta$ and $\etabar$ superfields were kept decouplingly heavy at 4 TeV. All the parameters were sampled over uniform prior distributions within their specific ranges and no biases were applied on any subset of the scanned parameters. The numerical scan was executed using our own code written in Python. All the constraints incorporating the respective mixing elements and branching ratios were built into this script.

\section{Results}\label{sec:result}

\begin{figure}[p]
\centering
\subfigure[\label{fig:parneut}]{\includegraphics[width=7.5cm]{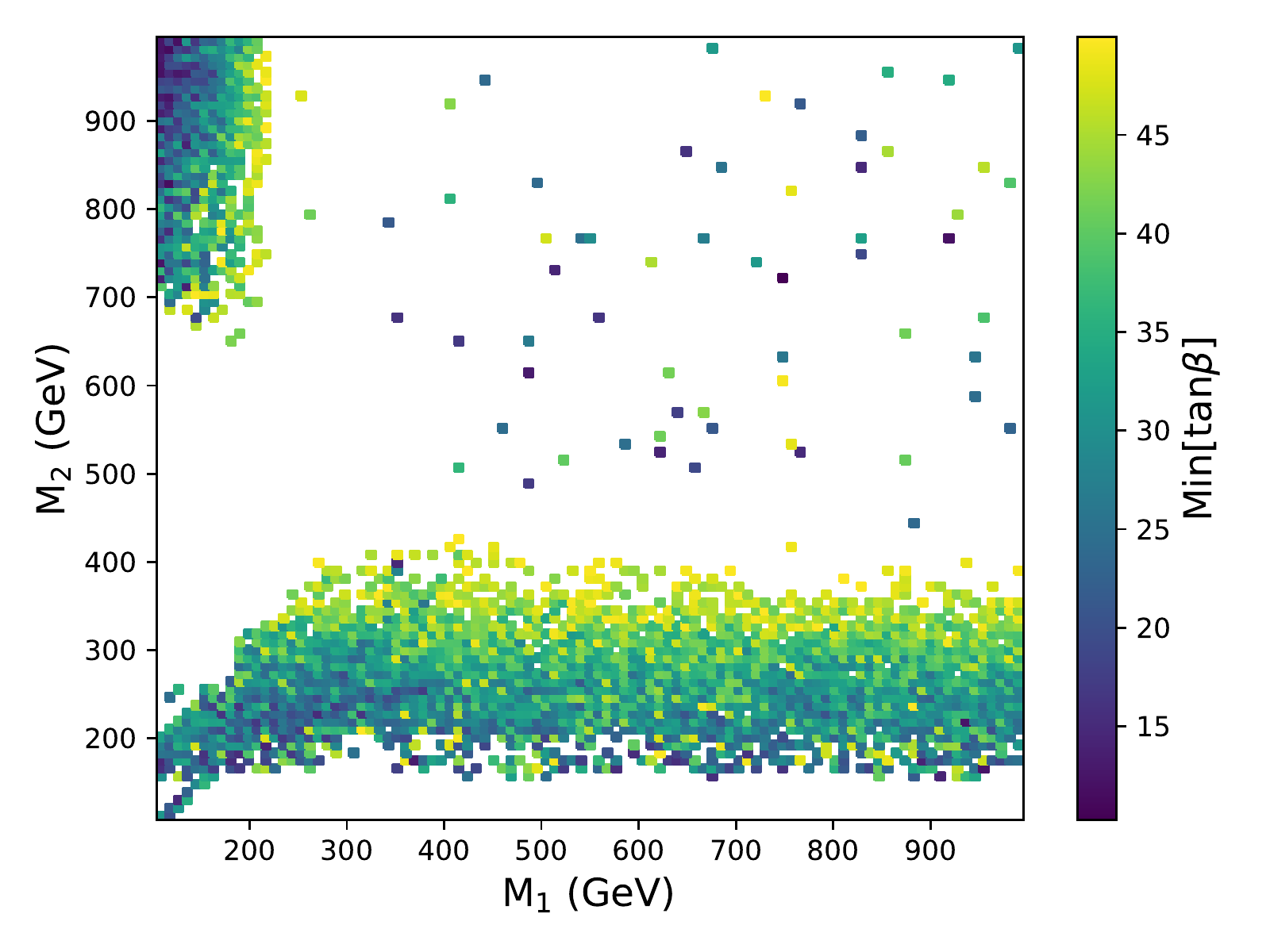}}
\subfigure[\label{fig:parcharged}]{\includegraphics[width=7.5cm]{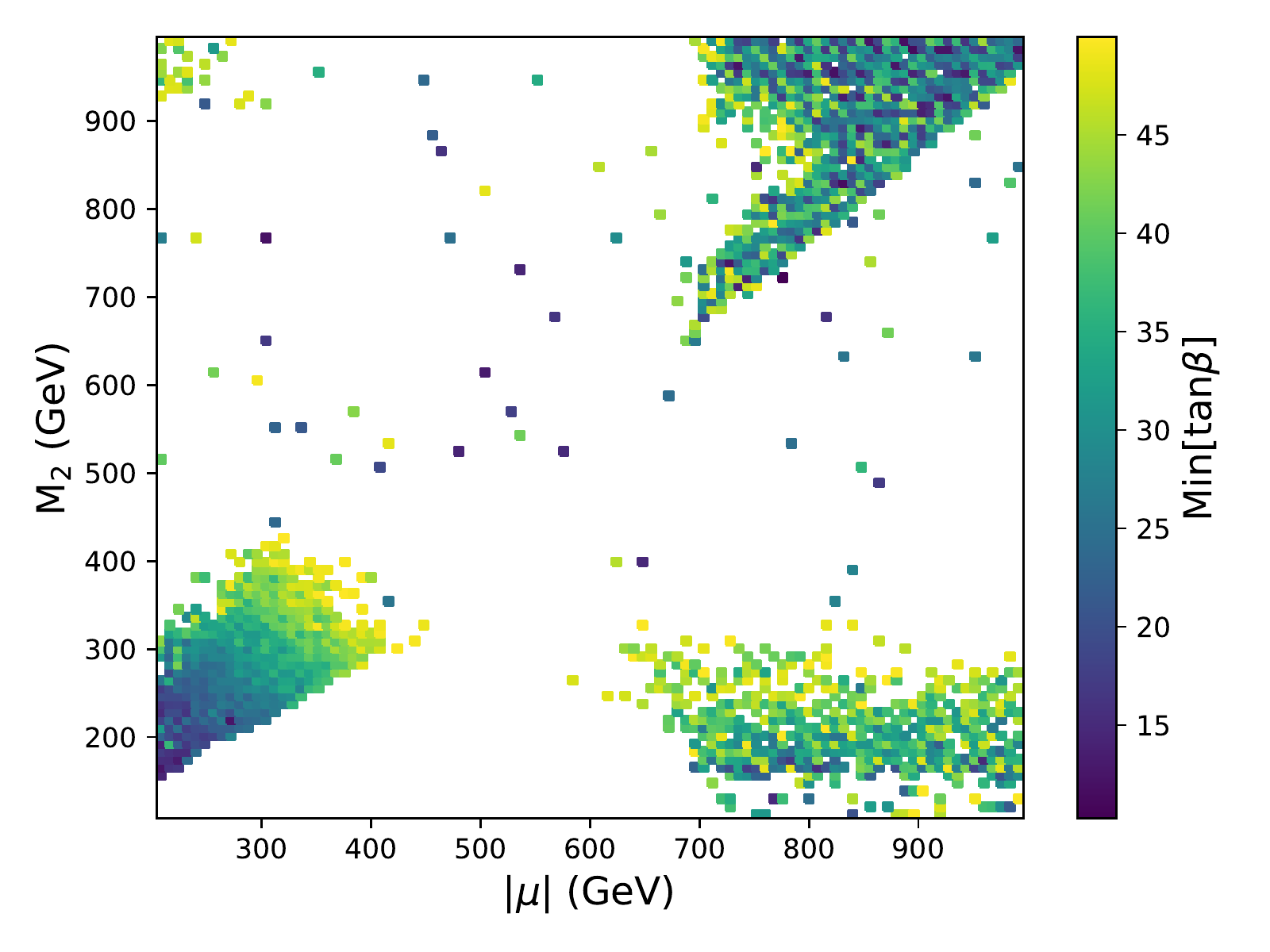}}
\subfigure[\label{fig:selchi}]{\includegraphics[width=7.5cm]{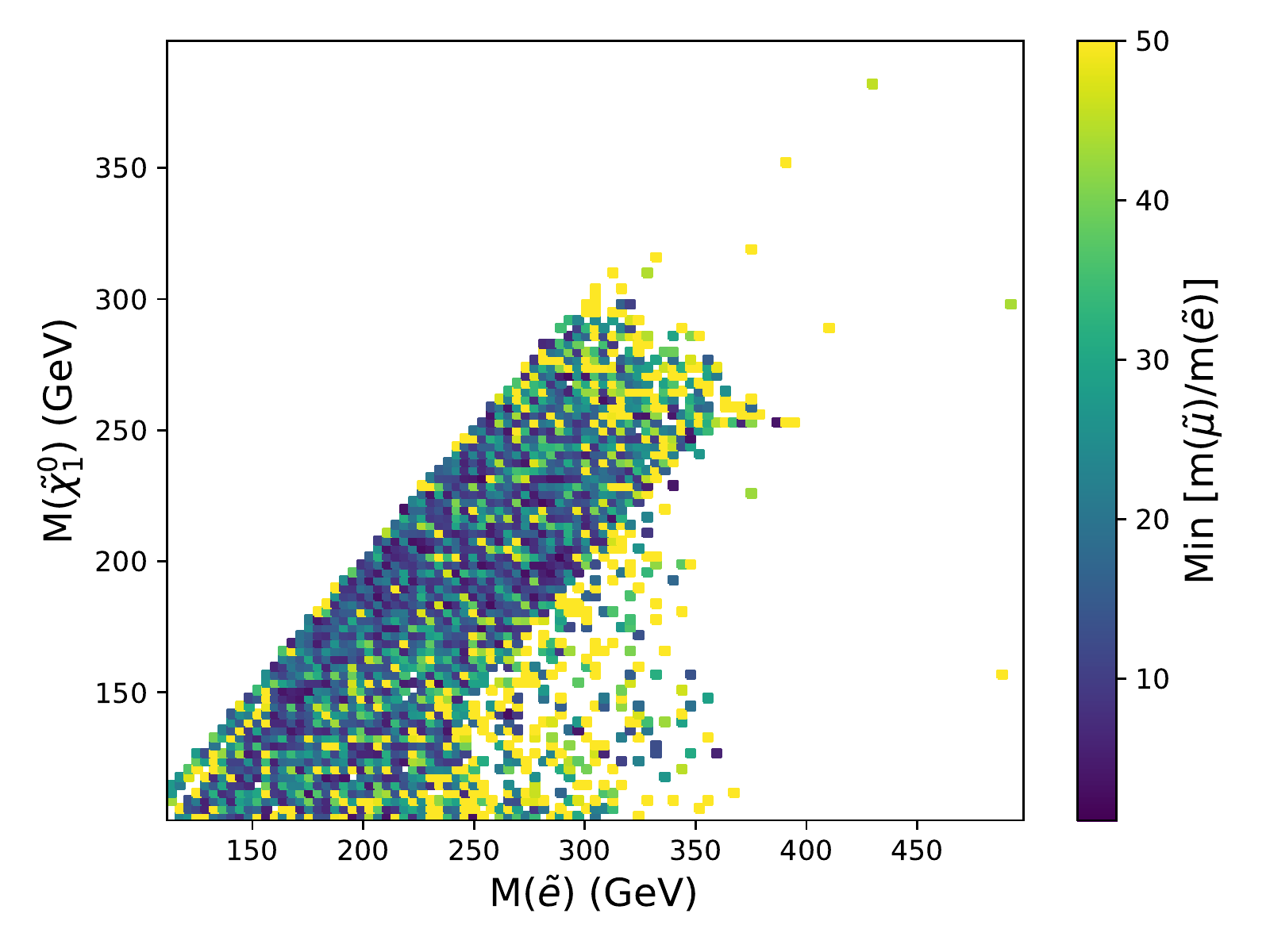}}
\subfigure[\label{fig:chachi}]{\includegraphics[width=7.5cm]{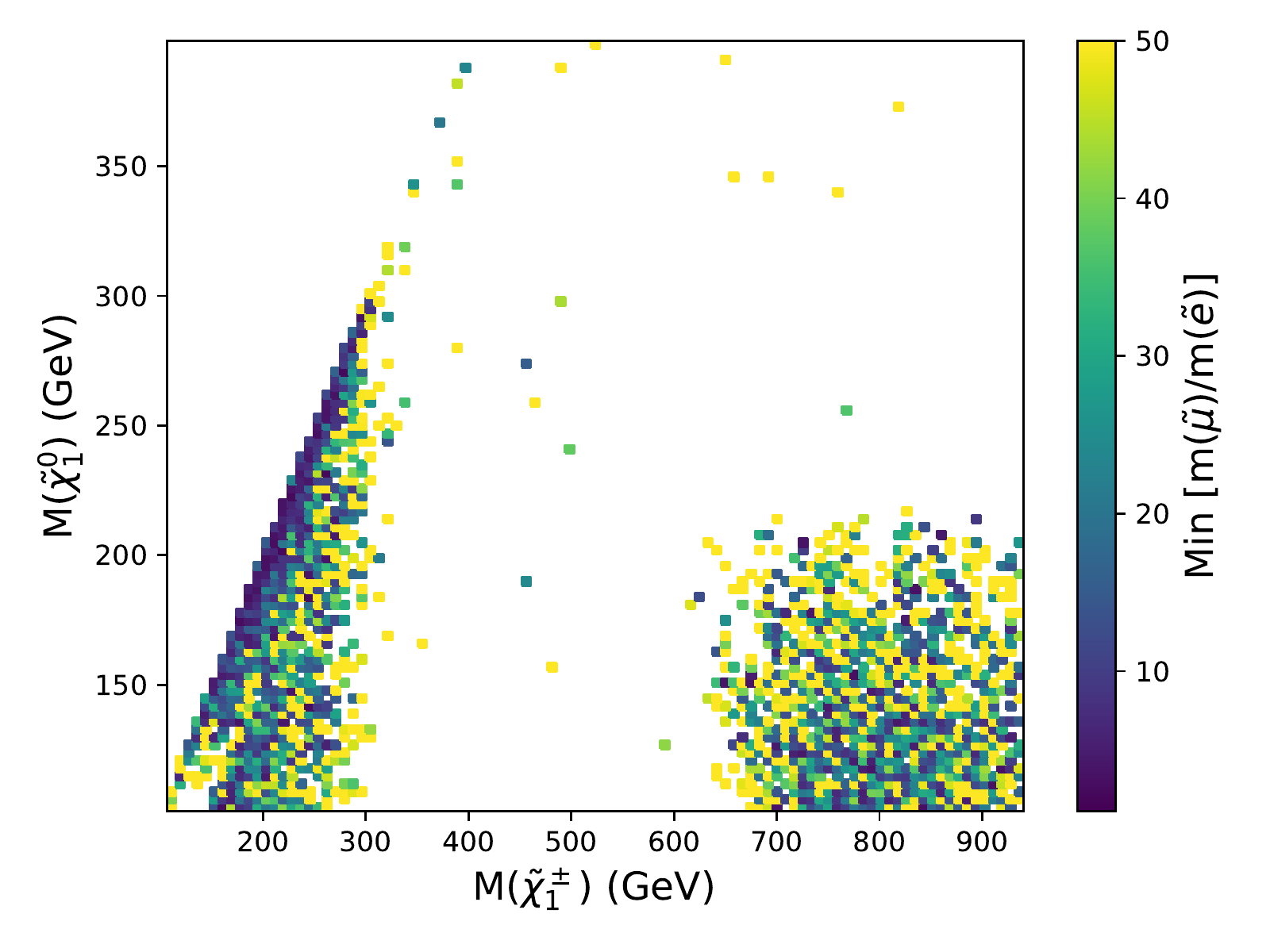}}
\subfigure[\label{fig:selratio}]{\includegraphics[width=7.5cm]{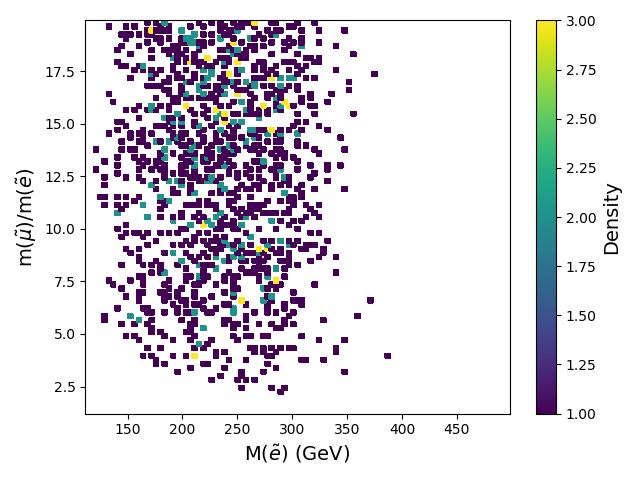}}
\subfigure[\label{fig:nlspdeltam}]{\includegraphics[width=7.5cm]{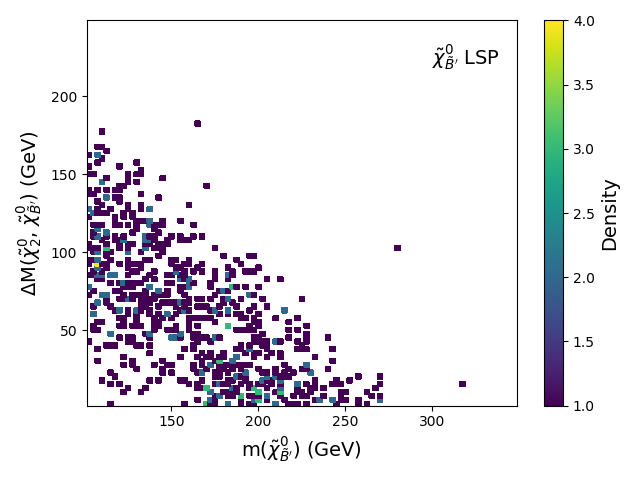}}
\caption{Parameter regions explaining both electron and muon anomalous magnetic moment observations: \ref{fig:parneut} shows the allowed regions in $M_1$-$M_2$ bins with the colors denoting minimum $\tan\beta$ encountered in the respective bins. Fig.\ref{fig:parcharged} plots the allowed bins of $\vert\mu\vert$ and $M_2$ with the same attribute for the color. \ref{fig:chachi}and \ref{fig:selchi} plot the viable spectrum of the lightest neutralino, chargino and slepton masses with the minimum encountered smuon to selectron mass ratio denoted by the colors. \ref{fig:selratio} shows the allowed selectron masses against the ratio of smuon to selectron masses, the colors denote the density of points. \ref{fig:nlspdeltam} shows parameter points where the LSP is $\bpino$ dominated against the mass gap between it and the NLSP.\label{fig:md-meu}}
\end{figure}

The results of the numerical scan are illustrated in Figure \ref{fig:md-meu}. Figures \ref{fig:parneut} and \ref{fig:parcharged} describe the allowed region of the $M_1$, $M_2$, $\vert\mu\vert$ and $\tan\beta$ parameter space. The figures are binned in the gaugino and higgsino mass parameters while the color code signifies the minimum encountered value of $\tan\beta$ for each bin. Figures \ref{fig:selchi} and \ref{fig:chachi} show the allowed values of the lightest physical neutralino, chargino and selectron masses. They are binned in the $\tilde{\chi}^0_1$, $\tilde{\chi}^{\pm}_1$ and $\tilde{e}_{\rm L/R}$ masses. The color code in these figures denote the minimum encountered value of the ratio of second to first generation slepton masses in each bin. White regions are bins where our scan could not find parameter choices that can explain the $(g-2)$ observations or are excluded by any of the constraints explained in Section \ref{sec:const}. The bin widths in all the figures are adjusted so that we have 100 bins along each axis.

\begin{table}[t]
\centering
\begin{tabular}{|c|c|c|c|c|c|}
\hline\hline
&&&&&\\
Label & M$_{\tilde{\chi}^0_1}$ (GeV) & M$_{\tilde{e}}$ (GeV) & M$_{\tilde{\mu}}$ (GeV) & M$_{Z^\prime}$ (MeV) & g$_{X}$/$10^{-4}$ \\
&&&&&\\
\hline
&&&&&\\
Allowed Range & $< 400$ & $< 385$ & $ > 605$ & $10- 210$ & $3.76 - 10$\\
&&&&&\\
\hline
\hline
\end{tabular}
\caption{Allowed physical neutralino, selectron and smuon masses as well as additional gauge boson mass and coupling that can explain both electron and muon $(g-2)$. Unlisted physical states/ parameters remain viable throughout their scanned ranges.}\label{table:allowed}
\end{table}

Figures \ref{fig:selchi} and \ref{fig:chachi} illustrate the upper bounds on the selectron and the lightest neutralino masses listed in Table \ref{table:allowed}. These upper limits come from the necessity to explain the $(g-2)_e$ that requires these fields to be light enough. In addition, there are gaps in the allowed lightest chargino spectrum which arises due to a combination of both LHC and $(g-2)$ observations. Heavier chargino masses require the neutralino to be lighter so as to explain $(g-2)_e$. However, a larger mass gap between these two fields beget strong exclusions from collider observations that study chargino decay modes into leptons and the lightest neutralino. These constraints restrict chargino masses unless they are heavy enough that their production cross section falls below the corresponding upper limits. Parameter regions with such large values of chargino masses require light selectrons and neutralinos to be able to explain the electron magnetic dipole moment.

In the light of these observations, we can classify the allowed parameter space into four broad categories.
\begin{enumerate}
    \item Light wino light higgsino: This region is represented by BP1 in Table \ref{table:bp}. The LSP is either wino-like or a wino-higgsino admixture. The lightest chargino can be nearly degenerate with the wino-like LSP which might make it stable enough to show up in long-lived charged particle searches at the LHC. The presence of a large number of light fields in a mildly compressed scenario makes it relatively easy to satisfy $(g-2)_e$. The same feature makes it difficult to compensate for the negative SUSY contribution to $(g-2)_\mu$. This allows for a relatively low value of $\tan\beta$. Apart from constraints on long-lived charginos, bounds from the LHC are weak for this choice of parameters because of the compressed nature of the lighter side of the spectrum.
    \item Light wino heavy higgsino: Represented by BP2 in Table \ref{table:bp}, this class of parameter choice is marked by a light wino-like neutralino and lightest chargino. The LSP is mostly bino-like which, along with comparatively larger values of $\tan\beta$, is necessitated by the need to explain the $(g-2)$ observations. LHC constraints on the light wino and the bino-like LSP may be severe depending on slepton masses. This forces the LSP, NLSP and the lightest chargino to be within $\lesssim 100$ GeV of one another.
    \item Heavy wino light higgsino: This class of allowed parameter choice is marked by a light higgsino triplet, one of which may also be the LSP. While the constraints on such a scenario are very weak indeed, it requires a very light bino-like neutralino and quite large values of $\tan\beta$ to be able to explain the lepton $(g-2)$. The LSP is often a bino-higgsino mixed state but may as well be either bino or higgsino dominated. This is represented by the BP3 in Table \ref{table:bp}.
    \item Heavy wino heavy higgsino: BP4 in Table \ref{table:bp} illustrates a sample point for this choice of parameters. This class of parameters would be able to bypass most LHC constraints as most of the fields remain quite heavy. The $(g-2)$ requirements are met entirely by a light Bino-like LSP and a very light selectron even with $\tan\beta$ as low as 20. However, this scenario requires the mass gap between these two fields to be small enough to have evaded detection at the LHC. 
\end{enumerate}

Figure \ref{fig:selratio} serves to illustrate a significant observation: a lower limit on the second generation slepton mass and an upper limit on the first generation slepton mass. It is binned in the selectron mass and the second to first generation slepton mass hierarchy. The color code shows the number of hits in that bin from the parameter scan. The visible lower limit on the slepton mass hierarchy translates to a corresponding lower limit on the second generation slepton mass as listed in Table \ref{table:allowed}. As mentioned earlier, the upper limit on selectron masses come from the $(g-2)_e$ observations. The lower limit on the smuon masses arise due to the necessity to suppress the negative SUSY contribution to $(g-2)_\mu$.

Figure \ref{fig:nlspdeltam} is binned in the mass of the LSP and the mass gap between it and the NLSP. However, this figure considers only the cases when the LSP is $\bpino$ dominated, which also means that the NLSP is necessarily an MSSM-like neutralino. In this scenario, the NLSP sets the scale for the rest of the particle spectrum. The need to explain the electron $(g-2)$ thus sets an upper limit on the NLSP-LSP mass gap.
\begin{table}[t]
\centering
\resizebox{\textwidth}{!}{%
\begin{tabular}{|c|c|c|c|c|c|c|c|c|c|c|c|c|c|c|c|c|}
\hline\hline
Label & M$_0$& M$_1$& M$_2$& $\vert\mu\vert$& $\tan\beta$ & M$_{\tilde{\chi}^0_1}$& M$_{\tilde{\chi}^0_2}$& M$_{\tilde{\chi}^0_3}$& M$_{\tilde{\chi}^0_4}$& M$_{\tilde{\chi}^0_5}$& M$_{\tilde{\chi}^{\pm}_1}$& M$_{\tilde{\chi}^{\pm}_2}$& M$_{\tilde{e}}$& M$_{\tilde{\mu}}$ & M$_{Z^\prime}$& g$_{X}$\\
& (GeV) & (GeV) & (GeV) & (GeV) & & (GeV) & (GeV) & (GeV) & (GeV) & (GeV) & (GeV) & (GeV) & (GeV) & (GeV) & (MeV) &$\times 10^{-4}$\\
\hline
&&&&&&&&&&&&&&&&\\
BP1 & 606 & 781 & 180 & 223 & 27 & 149 & 233 & 261 & 606 & 783 & 151 & 269 & 170 & 845 & 27 & 8.26\\
&&&&&&&&&&&&&&&&\\
\hline
&&&&&&&&&&&&&&&&\\
BP2 & 991 & 192 & 244 & 911 & 35 & 192 & 243 & 915 & 917 & 991 & 243 & 918 & 201 & 773 & 86 & 9.74\\
&&&&&&&&&&&&&&&&\\
\hline
&&&&&&&&&&&&&&&&\\
BP3 & 490 & 156 & 977 & 223 & 45 & 146 & 228 & 231 & 490 & 984 & 222 & 984 & 169 & 647 & 17 & 7.91\\
&&&&&&&&&&&&&&&&\\
\hline
&&&&&&&&&&&&&&&&\\
BP4 & 184 & 140 & 953 & 758 & 23 & 139 & 184 & 746 & 761 & 969 & 746 & 969 & 160 & 731 & 11 & 7.09\\
&&&&&&&&&&&&&&&&\\
\hline
\hline
\end{tabular}}
\caption{Sample benchmark points representing four different classes of allowed parameter choices.}\label{table:bp}
\end{table}

\section{Conclusion} \label{sec:conclusion}

We found that it is possible to explain the discrepancies between the theoretical and experimental values of both the electron and muon magnetic moments simultaneously in a supersymmetric gauged \newu model. Not only is it possible, but it is rather simple to obtain a viable parameter space bypassing even the most conservative of bounds from present SUSY searches. A large variety of parameter choices with a possible mass spectrum spanning the entire sub-TeV range may be able to explain the lepton magnetic moments in tandem. In spite of imposing no particular hierarchy on the parameters scanned numerically, a clear hierarchy is observed between the first and second generation sleptons. The first generation sleptons are required to be quite light ($\lesssim 400$ GeV), yet, unlike models explaining both lepton MDMs with modifications of MSSM parameters, they are not limited to be as light as to be barely above the old LEP bounds. Neither do we require tremendously large values of $\tan\beta$ or the $\mu$-parameter in order to satisfy the $(g-2)_e$.  In fact, viable choices of the parameter space include values of $\tan\beta$ as low as 20 and scenarios where the higgsinos form a light, nearly degenerate triplet at the lower end of the spectrum. There are gaps in the $\tilde{\chi}^{\pm}_1$ spectrum which are ruled out by a combination of $(g-2)$ requirements and collider results. Satisfying the $(g-2)$ observations in scenarios with the $\tilde{\chi}^{\pm}_1$ mass exceeding $\sim$300 GeV requires the $\tilde{\chi}^{0}_1$ to be light enough that they get excluded by LHC results. These bounds can be evaded when $\tilde{\chi}^{\pm}_1$ masses are greater than about 650 GeV.

In contrast to supersymmetric models explaining the lepton $(g-2)$ by introducing flavor-violating SUSY-breaking terms, our model does not predict any charged-lepton flavor violation. The mystery of the lepton magnetic moments that had us bewildered by violating both the natural sign and mass hierarchies amongst themselves seems almost naturally solvable when we embed the gauged $L_{\mu}-L_{\tau}$ symmetry in a SUSY framework. The premise becomes even more exciting when we consider the fact that a very narrow slice of the additional gauge boson mass-coupling parameter plane is actually used to explain the MDMs. This narrow slice could  be explored by the ongoing COHERENT and CCM experiments and by the future experiments, e.g.,  DUNE\cite{Bakhti:2018avv,Altmannshofer:2019zhy,Ballett:2019xoj}, BELLE-II\cite{Cecchi:2020lja,Chen:2017cic,Araki:2017wyg,Banerjee:2018mnw,Adachi:2019otg} etc. 

Current projections for the High-Luminosity LHC\cite{CidVidal:2018eel} show that the $Wh$ channel with a $\mathit{l}+b\bar{b} + {\rm MET}$ final state may be the strongest probe of the electroweakinos\cite{Chakraborti:2020vjp}, however it still cannot probe many of the scenarios described in this work. Especially, the parameter choices involving mildly compressed spectra amongst the lightest electroweakinos and sleptons would remain unconstrained. The possibility to explain both lepton $(g-2)$ with heavy charginos would be restricted as the lower limits on their masses are expected to rise in case there are no significant observations by the HL-LHC. The absence of projections for slepton or sneutrino mediated decays of charginos or neutralinos pair produced in $pp$ collisions stands out in this regard, as does that of direct selectron/smuon production, which lay the strongest exclusions on their respective particle masses from the latest LHC 13 TeV data\cite{Aaboud:2018jiw, Aad:2019vnb}. A larger analysis that can extend the HL-LHC projections from $3\mathit{l}+{\rm MET}$ final states probes via $WW/WZ$ mediated decays of the pair produced charginos/neutralinos to sub-500 GeV $\tilde{\chi}^\pm_1$ masses is key to understanding whether we will be able to probe the proposed scheme at the Large Hadron Collider in the recent future. We leave this alongside projections from considering direct production of selectron/smuon pairs to a future work.

\acknowledgments{
HB thanks Sabyasachi Chakraborty and Subhadeep Mondal for helpful discussions and Shu Liao for the introduction to Python. We also thank A. Thompson, L Strigari, G. Rich and D. Pershey for very useful discussions regarding the analysis of the recently released COHERENT LAr  data.  The work of BD is supported in part by the DOE Grant No. DE-SC0010813. SR acknowledges the hospitality of the Mitchell Institute for Fundamental Physics and Astronomy where part of this work was carried out.}

%\bibliography{refs2}

\providecommand{\href}[2]{#2}\begingroup\raggedright\endgroup

\end{document}